\documentclass[12pt,oneside]{article}
\usepackage[T1]{fontenc}
\title{Diagonal Matrix Elements of the Effective Hamiltonian
for $K^{0}-\bar{K}^{0} $ System \\
in One Pole Approximation}
\author{J. Jankiewicz\\
\small{University of Zielona G\'ora, Institute of Physics}\\
\small{Podg\'orna 50, 65-246 Zielona G\'ora, Poland}\\
\small{email: jjank@proton.if.uz.zgora.pl}}
\begin{document}
\bibliographystyle{plain}
\maketitle
\begin{abstract}
We study the properties of time evolution of the
$K^{0}-\bar{K}^{0} $ system in spectral formulation. Within the
one--pole model we find the exact form of the diagonal matrix
elements of the effective Hamiltonian for this system. It appears
that, contrary to the Lee--Oehme--Yang (LOY) result, these exact
diagonal matrix elements are different if the total system is
CPT--invariant but CP--noninvariant.
\end{abstract}

\section{Introduction}
This paper has been inspired by the results presented in \cite{p:1}
and \cite{p:2}. Paper \cite{p:1} analyses the problem of equality of
particle and antiparticle masses, whereas \cite{p:2} describes an
exactly solvable model of the particle-antiparticle system -- in
this particular case $K^{0}-\bar{K}^{0}$. The most important
properties of antiparticles follow from the CPT symmetry. This
symmetry, also known as the CPT theorem \cite{p:3}, determines the
properties of the transition amplitudes under the action of the
product of C, P and T transformations (charge conjugation, space
inversion and time reversal, respectively). According to the CPT
theorem, the transition amplitudes describing any physical process
must be CPT-invariant. From this we conclude that the full
Hamiltonian $H $ of the system under consideration must be invariant
under the product of the $\mathcal{C} $, $\mathcal{P} $ and
$\mathcal{T} $ operators. Another conclusion that can be drawn here
is that stable particles and their antiparticles must have the same
mass. This property of the particle-antiparticle pair is true for
stable particles and the same is usually assumed of unstable
particles (e.g. $K^{0}$ and $\bar{K}^{0} $). Such an extension of a
law true for stable particles to unstable particles is questioned in
\cite{p:1}. The reason for the widespread belief that this extension
is correct is most probably the Lee, Oehme and Yang (LOY)
approximation and the conclusions which follow from it --- and more
specifically, the properties of the effective Hamiltonian, $H_{LOY},
$ governing the time evolution in the subspace
$\mathcal{H}_{\parallel}.$ In our case $\mathcal{H}_{\parallel}$ is
the subspace of the total Hilbert space of states $\mathcal{H}$,
spanned by state vectors of $K^{0},$ $\bar{K}^{0} $ mesons.

Following the LOY approach, a nonhermitian Hamiltonian
$H_{\parallel} $ is usually used to study the properties of the
particle-antiparticle unstable system \cite{p:5}-\cite{p:11}
\begin{equation}
 H_{\parallel}\equiv M-\frac{i}{2}\Gamma, \label{j1-5}
\end{equation}
where
\begin{equation}
M=M^{+} \ , \  \Gamma = \Gamma^{+}  \label{j1-6}
\end{equation}
are $(2\times2) $ matrices acting in $\mathcal{H}_{\parallel}$. The
$M $-matrix is called the mass matrix and $\Gamma$ is the decay
matrix. Lee, Oehme and Yang derived their approximate effective
Hamiltonian $H_{\parallel}\equiv H_{LOY} $ by adapting the
one-dimensional Weisskopf-Wigner (WW) method to the two-dimensional
case corresponding to the neutral kaon system. Almost all properties
of this system can be described by solving the Schr\"{o}dinger--like
equation
\begin{equation}
i\frac{\partial }{\partial t}|\psi
;t\rangle_{\parallel}=H_{\parallel }|\psi ;t\rangle_{\parallel}, \ \
\  (t\geq t_{0}> -\infty ) \label{j1-7}
\end{equation}
(where we have used  $\hbar =c=1 $). The initial conditions for Eq.
(\ref{j1-7}) are
\begin{equation}
\parallel |\psi ;t=t_{0}\rangle_{\parallel} \parallel =1,
\ \ |\psi ;t_{0}=0\rangle_{\parallel} =0, \label{j1-8}
\end{equation}
where  $|\psi ;t=t_{0}\rangle_{\parallel} $ belongs to
$\mathcal{H}_{\parallel} $ ($\mathcal{H}_{\parallel}\subset
\mathcal{H} $) and $\mathcal{H}_{\parallel}$ is  spanned by
orthonormal neutral kaons states: $|K^{0}\rangle  \equiv
|\textbf{1}\rangle $, $|\bar{K}^{0}\rangle \equiv
|\textbf{2}\rangle$. Thus $\mathcal{H}_{\parallel} = P\mathcal{H} $,
where
\begin{equation}
P\equiv |\textbf{1}\rangle \langle\textbf{1}|+ |\textbf{2}\rangle
\langle\textbf{2}|. \label{j1-18}
\end{equation}

According to the  standard result of the LOY approach, in a CPT
invariant system, i.e. when
\begin{equation}
\Theta H \Theta^{-1}=H, \label{j1-3}
\end{equation}
(where $\Theta = CPT$) we have
\begin{equation}
h_{11}^{LOY}=h_{22}^{LOY} \label{j1-14}
\end{equation}
and
\begin{equation}
M_{11}^{LOY}=M_{22}^{LOY}, \label{j1-15}
\end{equation}
where: $M_{jj}^{LOY}=\Re (h_{jj}^{LOY}) $ and $\Re (z)$ denotes the
real part of a complex number $z$ (then $\Im (z) $ is the imaginary
part of $z$), and
 $h_{jj}^{LOY}=\langle \textbf{j}|H_{LOY} |\textbf{j} \rangle $
 $(j=1,2) $.

The universal properties of the two particles subsystem described by
the $H$ fulfilling the condition (\ref{j1-3}), may be investigated
by the use of the matrix elements of the exact evolution operator
for ${\cal H}_{||}$ instead of the approximate one used in the LOY
theory. This exact evolution operator, $U_{\parallel}(t)$, can be
written as follows $U_{\parallel}(t)=PU(t)P$, where $U(t) \equiv
e^{\textstyle - itH} $ is the exact evolution operator acting in the
total state space $\cal H$.

Assuming that the CPT symmetry is conserved in the system under
considerations one finds that the matrix elements
\begin{equation}
A_{jk}(t)=\langle \textbf{j}|U_{\parallel}(t)|\textbf{k}\rangle
\equiv \langle \textbf{j}|U(t)|\textbf{k}\rangle \ \ (j,k=1,2),
\label{j1-23}
\end{equation}
of the exact  $U_{\parallel}(t)$, obey
\begin{equation}
A_{11}(t)=A_{22}(t). \label{j1-25}
\end{equation}

General conclusions concerning the properties of the difference of
the diagonal matrix elements $(h_{11} - h_{22})$ of the exact
$H_{\parallel}$, (which can in general depend on time $t$
\cite{h(t)}), where
\begin{equation}
h_{jk}=\langle \textbf{j}|H_{\parallel}|\textbf{k} \rangle
\,\,\,\,\, (j,k=1,2), \label{j1-12}
\end{equation}
can be drawn by analyzing the following expression derived in
\cite{p:1} for CPT invariant systems
\begin{equation}
h_{11}(t)-h_{22}(t)\equiv
\frac{i}{det\textbf{A}(t)}\Biggl(\frac{\partial A_{21}(t) }{\partial
t}A_{12}(t)-\frac{\partial A_{12}(t) }{\partial t}A_{21}(t)\Biggl).
\label{j1-40}
\end{equation}
In \cite{p:1} it is shown that
\begin{equation}
h_{11}(t)-h_{22}(t)\neq 0, \label{j1-46}
\end{equation}
when (\ref{j1-3}) holds and
\begin{equation}
[\mathcal{CP},H]\neq 0,  \label{j1-45}
\end{equation}
that is in the exact quantum theory the difference
$(h_{11}(t)-h_{22}(t))$  cannot be equal to zero with CPT conserved
and CP violated. In Section 3 we will consider this relation in the
context of an exactly solvable model. This problem is important
because realistic calculations are carried out with the use of
simplified and approximate models. Not all of them conform to the
requirements of the exact (not approximate) quantum theory.

The aim of this paper is to calculate the difference of the diagonal
matrix elements of the effective Hamiltonian, (\ref{j1-40}), in a
CPT invariant and CP noninvariant system for the approximate model
analyzed in \cite {p:2}, that is  in the case of the one -- pole
model based on the Breit -- Wigner ansatz, i.e. the same model as
used in Lee, Oehme and Yang theory.

The paper is organized as follows.  In Section 2 we review briefly
the spectral formulation for the neutral kaon system and a model
described in \cite{p:2}: one pole approximation. Section 3
investigates the diagonal matrix elements of the effective
Hamiltonian and their difference in the CPT invariant and CP
noninvariant system in the case of the one -- pole model. In Section
4 we present our conclusions and we estimate the numerical result of
the investigated difference for the $K^{0}-\bar{K}^{0} $ system.
Appendix A contains the relevant integrals and derivatives used in
Section 3. In Appendix B we give the exact formulae for expressions
appearing in Section 3. In Appendix C we find coefficients which
were used in Section 3.

\section{The model: one pole approximation}
While describing the two and three pion decay we are mostly
interested in the $|K_{S}\rangle $ and $|K_{L}\rangle $
superpositions of $|K^{0}\rangle$ and $|\bar{K}^{0}\rangle.$ These
states correspond to the physical $|K_{S}\rangle $ and
$|K_{L}\rangle $ neutral kaon states \cite{p:2,p:25,p:26}
\begin{eqnarray}
|K_{S}\rangle =p|K^{0}\rangle +
q|\bar{K}^{0}\rangle, \ \ |K_{L}\rangle =p|K^{0}\rangle
-q|\bar{K}^{0}\rangle . \label{j1-47}
\end{eqnarray}
We assume that these physical states are the initial physical states
of the CPT -- invariant system, i.e. at the instant of creation of
neutral kaons. We have
\begin{eqnarray}
\langle K_{S}|K_{S}\rangle = \langle K_{L}|K_{L}\rangle \equiv
|p|^{2}+|q|^{2}=1 \label{j1-48}
\end{eqnarray}
\begin{eqnarray}
\langle K_{S}|K_{L}\rangle = \langle K_{L}|K_{S}\rangle \equiv
|p|^{2}-|q|^{2} \stackrel{def}{=} \Delta_{K} \neq 0. \label{j1-49}
\end{eqnarray}

The time evolution of $K^{0} $ and $\bar{K}^{0}$ can be concisely
presented in the following way:
\begin{eqnarray}
|K_{\alpha}(t)\rangle&=&e^{-iHt}|K_{\alpha}\rangle \nonumber \\
&\equiv&
A_{K_{\alpha}K_{\alpha}}(t)|K_{\alpha}\rangle+
A_{K_{\alpha}K_{\beta}}(t)|K_{\beta}\rangle
+Q e^{-iHt}|K_{\alpha}\rangle ,  \label{j1-50}
\end{eqnarray}
where $K_{\alpha}=K^{0},  \bar{K}^{0}$ and $H $ is the full
hermitian Hamiltonian and $Q=\textit{I}-P$,
\begin{eqnarray}
A_{K_{\alpha}K_{\beta}}(t) = \langle
K_{\alpha}|e^{-iHt}|K_{\beta}\rangle \equiv \langle
K_{\alpha}|K_{\beta}(t)\rangle .  \label{j1-50b}
\end{eqnarray}

Let us notice that amplitudes $A_{K_{\alpha}K_{\beta}}(t)$ for
$K_{\alpha}, K_{\beta}=K^{0}, \bar{K}^{0}$ correspond to the
previously defined amplitudes\footnote{Amplitudes
$A_{K^{0}K^{0}}(t),$ etc., correspond to $P_{K^{0}K^{0}}(t),$ etc.
respectively used in \cite{p:2}} $A_{jk}(t), $ where $j, k = 1, 2$,
(\ref{j1-23}). Consequently we may write
\begin{eqnarray}
\nonumber&& A_{K^{0}K^{0}}(t)\equiv \langle K^{0}|e^{-itH}
|K^{0}\rangle= \langle \textbf{1}|e^{-itH}
|\textbf{1}\rangle\equiv A_{11}(t),
\nonumber\\
&& A_{K^{0}\bar{K}^{0}}(t)\equiv \langle K^{0}|e^{-itH}
|\bar{K}^{0}\rangle= \langle \textbf{1}|e^{-itH}
|\textbf{2}\rangle\equiv A_{12}(t),
\nonumber\\
&& A_{\bar{K}^{0}K^{0}}(t)\equiv \langle \bar{K}^{0}|e^{-itH}
|K^{0}\rangle= \langle \textbf{2}|e^{-itH}
|\textbf{1}\rangle\equiv A_{21}(t),
\nonumber\\
&& A_{\bar{K}^{0}\bar{K}^{0}}(t)\equiv \langle
\bar{K}^{0}|e^{-itH} |\bar{K}^{0}\rangle= \langle
\textbf{2}|e^{-itH} |\textbf{2}\rangle\equiv A_{22}(t).
\label{j1-50a}
\end{eqnarray}
Using the spectral formalism we can write unstable states $|\lambda
\rangle $ as
\begin{eqnarray}
|\lambda \rangle =\sum_{q} \omega_{\lambda}(q)|q \rangle
\label{j1-62-a}
\end{eqnarray}
and then $|\lambda (t)\rangle $ as
\begin{eqnarray}
|\lambda (t)\rangle \stackrel{def}{=}e^{-i t H} |\lambda \rangle
=\sum_{q}|q (t)\rangle \omega_{\lambda}(q), \label{j1-62}
\end{eqnarray}
where $|q (t)\rangle =e^{-itH}|q \rangle $ and vectors $|q \rangle $
form a complete set of eigenvectors of the hermitian,
quantum-mechanical Hamiltonian $H$ and $ \omega_{\lambda}(q) =
\langle q|\lambda \rangle.$ If the continuous eigenvalue is denoted
by $m$, we can define the survival amplitude  $A(t)$ (or the
transition amplitude in the case of $K^{0}\leftrightarrow
\bar{K}^{0} $ ) in the following way:
\begin{eqnarray}
A(t)=\int\limits_{Spec(H)}dm \; e^{-imt}\rho (m), \label{j1-62a}
\end{eqnarray}
where the integral extends over the whole spectrum of the
Hamiltonian and density $\rho (m)$ is defined as follows
\begin{eqnarray}
\rho (m)=|\omega_{\lambda}(m)|^{2}, \label{j1-62b}
\end{eqnarray}
where $\omega_{\lambda}(m)=\langle m |\lambda \rangle . $

The above formalism may be applied to  $|K_{S}\rangle $ and
$|K_{L}\rangle $ by introducing a hermitian Hamiltonian with a
continuous spectrum of decay products labelled by  $\alpha, \beta
$ etc.,
\begin{eqnarray}
H|\phi_{\alpha}(m)\rangle =m \, |\phi_{\alpha}(m)\rangle , \
\langle \phi_{\beta}(m')|\phi_{\alpha}(m) \rangle = \delta_{\alpha
\beta }\delta (m'-m). \label{j1-65}
\end{eqnarray}
In accordance with formula (\ref{j1-62}) the unstable states $K_{S}
$ and $K_{L} $ may now be written as a superposition of the
eigenkets $|\phi_{\alpha}(m) \rangle ,$
\begin{eqnarray}
|K_{S}\rangle =\int_{0}^{\infty }dm \;
\sum_{\alpha}\omega_{S,\alpha}(m)|\phi_{\alpha}(m)\rangle ,
\label{j1-66}
\end{eqnarray}
\begin{eqnarray}
|K_{L}\rangle =\int_{0}^{\infty }dm \;
\sum_{\beta}\omega_{L,\beta}(m)|\phi_{\beta}(m)\rangle .
\label{j1-67}
\end{eqnarray}
Thus
\begin{eqnarray}
|K_{S}(t)\rangle =e^{-itH}|K_{S}\rangle =  \int_{0}^{\infty}dm\;
\sum_{\alpha}\omega_{S,\alpha }(m)e^{-itH}| \phi_{\alpha
}(m)\rangle . \label{j1-67a}
\end{eqnarray}
\begin{eqnarray}
|K_{L}(t)\rangle =e^{-itH}|K_{L}\rangle =  \int_{0}^{\infty}dm \;
\sum_{\beta}\omega_{L,\beta }(m)e^{-itH}| \phi_{\beta }(m)\rangle
. \label{j1-67b}
\end{eqnarray}
Using (\ref{j1-67a}) and (\ref{j1-67b}) we can write
\begin{eqnarray}
\nonumber&& \langle K_{S}|K_{S}(t)\rangle =\int_{0}^{\infty } dm \;
\sum_{\alpha}|\omega_{S,\alpha}(m)|^{2}e^{-imt}\stackrel{def}{=}
A_{K_{S}K_{S}}(t),\nonumber\\
&& \langle K_{L}|K_{L}(t)\rangle = \int_{0}^{\infty }dm \;
\sum_{\beta}|\omega_{L,\beta}(m)|^{2}e^{-imt}
\stackrel{def}{=}A_{K_{L}K_{L}}(t),\nonumber\\
&&\langle K_{S}|K_{L}(t)\rangle =\int_{0}^{\infty }dm \;
\sum_{\gamma}\omega_{S,\gamma}^{\ast}
(m)\omega_{L,\gamma}(m)e^{-imt}\stackrel{def}{=}
A_{K_{S}K_{L}}(t),\nonumber\\
&&\langle K_{L}|K_{S}(t)\rangle =\int_{0}^{\infty }dm \;
\sum_{\sigma}\omega_{L,\sigma}^{\ast}(m)\omega_{S,\sigma}
(m)e^{-imt}\stackrel{def}{=} A_{K_{L}K_{S}}(t). \label{j1-68}
\end{eqnarray}
From (\ref{j1-47}) we can obtain
\begin{eqnarray}
|K^{0}\rangle = \frac{1}{2p}(|K_{S}\rangle +|K_{L}\rangle
)\label{j1-68a}
\end{eqnarray}
and
\begin{eqnarray}
|\bar{K}^{0}\rangle = \frac{1}{2q}(|K_{S}\rangle
-|K_{L}\rangle).\label{j1-68b}
\end{eqnarray}

Now, using formulae (\ref{j1-50a}), (\ref{j1-67a})-(\ref{j1-68}),
we can express $A_{K^{0}K^{0}}(t) $ etc. in terms of quantities
describing physical states, that is through the amplitudes
$A_{K_{S}K_{S}}(t) $, $A_{K_{L}K_{S}}(t) $ etc. (see eg.
\cite{p:2})
\begin{eqnarray}
A_{K^{0}\bar{K}^{0}}(t)\equiv \frac{1}{4p^{\ast }q}\left[A_{K_{S}K_{S}}(t)-
A_{K_{L}K_{L}}(t)-A_{K_{S}K_{L}}(t)+A_{K_{L}K_{S}}(t)\right], \label{j1-58}
\end{eqnarray}
\begin{eqnarray}
A_{\bar{K}^{0}K^{0}}(t)\equiv \frac{1}{4p
q^{\ast}}\left[A_{K_{S}K_{S}}(t)-
A_{K_{L}K_{L}}(t)+A_{K_{S}K_{L}}(t)-A_{K_{L}K_{S}}(t)\right].
\label{j1-59}
\end{eqnarray}
One can also find that
\begin{eqnarray}
A_{K_{S}K_{S}}+A_{K_{L}K_{L}}=
2\,(|p|^{2}A_{K^{0}K^{0}}(t)+|q|^{2}A_{\bar{K}^{0}\bar{K}^{0}}(t)\,).
\label{j1-59a}
\end{eqnarray}
Assuming (\ref{j1-3}) and using (\ref{j1-48}), (\ref{j1-59a}) we
get
\begin{eqnarray}
A_{K^{0}K^{0}}(t)=A_{\bar{K}^{0}\bar{K}^{0}}(t)
\equiv\frac{1}{2}\left(A_{K_{S}K_{S}}(t)+A_{K_{L}K_{L}}(t)\right).
\label{j1-60}
\end{eqnarray}
It follows from (\ref{j1-68a}), (\ref{j1-68b}) and (\ref{j1-58}) --
(\ref{j1-60}) that the probabilities $A_{K^{0}K^{0}}(t) $ etc. can
be written in the following way:
\begin{eqnarray}
A_{K^{0}K^{0}}(t)\nonumber&=&
A_{\bar{K}^{0}\bar{K}^{0}}(t)=\int_{0}^{\infty}
dm \; \rho_{K^{0}K^{0}}(m)e^{-imt}\nonumber \\
&=&\frac{1}{2}\int_{0}^{\infty }dm \;
\sum_{\alpha}\left\{|\omega_{S,\alpha }(m)|^{2}+|\omega_{L,\alpha
}|^{2}(m)\right\}e^{-imt} \label{j1-69}
\end{eqnarray}
\begin{eqnarray}
A_{K^{0}\bar{K}^{0}}(t)\nonumber&=& \int_{0}^{\infty
}dm \; \rho_{K^{0}\bar{K}^{0}}(m)e^{-imt}\nonumber \\
&=&\frac{1}{4p^{\ast }q}\int_{0}^{\infty }dm \;
\sum_{\beta}\Biggl\{|\omega_{S,\beta }(m)|^{2}-|\omega_{L,\beta
}(m)|^{2}\nonumber\\
&&-\omega_{S,\beta }^{\ast }(m)\omega_{L,\beta
}(m)+\omega_{L,\beta }^{\ast}(m)\omega_{S,\beta }(m)
\Biggl\}e^{-imt} \label{j1-70}
\end{eqnarray}
\begin{eqnarray}
A_{\bar{K}^{0}K^{0}}(t)\nonumber&=& \int_{0}^{\infty
}dm \; \rho_{\bar{K}^{0}K^{0}}(m)e^{-imt}\nonumber \\
&=&\frac{1}{4pq^{\ast }}\int_{0}^{\infty }dm \;
\sum_{\beta}\Biggl\{|\omega_{S,\beta }(m)|^{2}-|\omega_{L,\beta
}(m)|^{2}\nonumber\\
&&+\omega_{S,\beta }^{\ast }(m)\omega_{L,\beta
}(m)-\omega_{L,\beta }^{\ast}(m)\omega_{S,\beta }(m)
\Biggl\}e^{-imt}. \label{j1-71}
\end{eqnarray}

The Breit-Wigner ansatz \cite {p:27}
\begin{eqnarray}
\rho_{BW}(m)=\frac{\Gamma}{2\pi
}\frac{1}{(m-m_{0})^{2}+\frac{\Gamma^{2}}{4}} \equiv |\omega
(m)|^{2} \label{j1-63}
\end{eqnarray}
leads to the well known exponential decay law which follows from
the survival amplitude
\begin{eqnarray}
A_{BW}(t)=\int_{-\infty}^{\infty}dm \;
e^{-imt}\rho_{BW}(m)=e^{-im_{0}t}e^{-\frac{1}{2}\Gamma|t|}.
\label{j1-64}
\end{eqnarray}
(Note that the existence of the lower bound for the energy (mass)
induces non-exponential corrections to the decay law and to the
survival amplitude (\ref{j1-64}) --- see \cite{p:2} ). It is
reasonable to assume a suitable form for $\omega_{S,\beta } $ and
$\omega_{L,\beta }$. More specifically, we use \cite{p:2}
\begin{equation}
\omega_{S,\beta}(m)= \sqrt{\frac{\Gamma_{S}}{2\pi}}
\frac{A_{S,\beta}(K_{S}\rightarrow\beta)}{m-m_{S}+
i\frac{\Gamma_{S}}{2}}, \label{j1-72}
\end{equation}
\begin{equation}
\omega_{L,\beta }(m)=\sqrt{\frac{\Gamma_{L}}{2\pi}}
\frac{A_{L,\beta}(K_{L}\rightarrow\beta)}{m-m_{L}+
i\frac{\Gamma_{L}}{2}} \label{j1-73}
\end{equation}
where  $A_{S,\beta} $ and  $A_{L,\beta} $ are the decay (transition)
amplitudes. It is convenient to use the following definitions:
\begin{equation}
\gamma_{S}\equiv \frac{\Gamma_{S}}{2}, \ \gamma_{L}\equiv
\frac{\Gamma_{L}}{2}, \  \Delta m\equiv m_{L}-m_{S} \label{j1-74}
\end{equation}
\begin{equation}
S\equiv \sum_{\alpha }|A_{S,\alpha }|^{2}, \  L\equiv \sum_{\alpha
}|A_{L,\alpha }|^{2} \label{j1-75}
\end{equation}
\begin{equation}
R \equiv \sum_{\sigma }\Re(A_{S,\sigma }^{\ast }A_{L,\sigma }), \
I \equiv \sum_{\sigma }\Im (A_{S,\sigma }^{\ast }A_{L,\sigma }).
\label{j1-76}
\end{equation}
In the one-pole approximation (\ref{j1-72}), (\ref{j1-73})
$A_{K^{0}K^{0}}(t) $ can be conveniently written as
\begin{eqnarray}
A_{K^{0}K^{0}}(t)&=&A_{\bar{K}^{0}\bar{K}^{0}}(t) \nonumber \\
&=&-\frac{1}{2\pi}\Biggl\{e^{-im_{S}t}
\left(-\int_{0}^{-\frac{m_{S}}{\gamma_{S}}}dy
\frac{e^{-i\gamma_{S}ty}}{y^{2}+1} +\int_{0}^{\infty}dy
\frac{e^{-i\gamma_{S}ty}}{y^{2}+1}\right)\nonumber\\
&&+e^{-im_{L}t} \left(-\int_{0}^{-\frac{m_{L}}{\gamma_{L}}}dy
\frac{e^{-i\gamma_{L}ty}}{y^{2}+1}+ \int_{0}^{\infty}dy
\frac{e^{-i\gamma_{L}ty}}{y^{2}+1}\right)\Biggl\}. \label{j1-77}
\end{eqnarray}
Collecting  only exponential terms in (\ref{j1-77}) one obtains an expression
analogous to the WW approximation
\begin{equation}
A_{K^{0}K^{0}}(t)=A_{\bar{K}^{0}\bar{K}^{0}}(t)=
\frac{1}{2}\left(e^{-im_{S}t}e^{-\gamma_{S}t}+
e^{-im_{L}t}e^{-\gamma_{L}t}\right)+N_{K^{0}K^{0}}(t).
\label{j1-78}
\end{equation}
Here $N_{K^{0}K^{0}}(t)$ denotes all non-oscillatory terms present
in the integrals (\ref{j1-77}).

\section{Diagonal matrix elements of the effective\\ Hamiltonian}
This section constitutes the main part of the paper. Using the
decomposition of type (\ref{j1-78}) and the one-pole ansatz
(\ref{j1-72}), (\ref{j1-73}), we find the difference (\ref{j1-46}),
which is now  formulated for the $K^{0} - \bar{K}^{0} $ system. It
can be written as follows
\begin{eqnarray}
h_{11}(t)-h_{22}(t)=\frac{X(t)}{Y(t)}, \label{j1-79}
\end{eqnarray}
where
\begin{eqnarray}
X(t)= i\left(\frac{\partial A_{\bar{K}^{0}K^{0}}(t)}{\partial
t}A_{K^{0}\bar{K}^{0}}(t)-\frac{\partial
A_{K^{0}\bar{K}^{0}}(t)}{\partial t}A_{\bar{K}^{0}K^{0}}(t)\right)
\label{j1-79a}
\end{eqnarray}
and
\begin{eqnarray}
Y(t)= A_{K^{0}K^{0}}(t)A_{\bar{K}^{0}\bar{K}^{0}}(t)
-A_{K^{0}\bar{K}^{0}}(t)A_{\bar{K}^{0}K^{0}}(t). \label{j1-79b}
\end{eqnarray}

To calculate (\ref{j1-70}), (\ref{j1-71}) we use the following
relations \cite{p:2}
\begin{eqnarray}
\int_{0}^{\infty}dm \; \sum_{\alpha}
|\omega_{S,\alpha}(m)|^{2}e^{-imt}\nonumber&=&
\frac{1}{\pi}e^{-im_{S}t} \Biggl[-J^{(0)}
(\gamma_{S}t,-\frac{m_{S}}{\gamma_{S}})\nonumber\\
&&+K^{(0)}(\gamma_{S}t)\Biggl] \label{j1-85}
\end{eqnarray}
and
\begin{eqnarray}
&&\int_{0}^{\infty} dm \, \sum_{\beta} \Im\left(\omega_{S,\beta}(m)
\, \varphi_{L,\beta}^{\ast}(m)\right)
\, \, e^{-imt}= \;\;\;\;\;\;\;\;\; \;\;\;\nonumber\\
&&=-\frac{\sqrt{\gamma_{S} \gamma_{L}}}{\pi}\int_{0}^{\infty}dm \;
\frac{a_{1}m^{2}+b_{1}m+ c_{1}}{[(m-m_{S})^{2}+
\gamma_{S}^{2}][(m-m_{L})^{2}+
\gamma_{L}^{2}]}e^{-imt} \nonumber\\
&&=-\frac{\sqrt{\gamma_{S}\gamma_{L}}}{\pi}
\Biggl\{\frac{e^{-im_{S}t}}{\gamma_{S}}
\Biggl(D'_{I}\left(-J^{(0)}(\gamma_{S}t,-
\frac{m_{S}}{\gamma_{S}})\right)+
K^{(0)}(\gamma_{S}t)\nonumber\\
&& \, \, \, \, \, \,  +\gamma_{S}C_{I}
\left(-J^{(1)}(\gamma_{S}t,-\frac{m_{S}}{\gamma_{S}})
\right)+K^{(1)}(\gamma_{S}t)\Biggl) \nonumber\\
&& \, \, \, \, \, \, +\frac{e^{-im_{L}t}}{\gamma_{L}}\Biggl(F'_{I}
\left(-J^{(0)}(\gamma_{L}t,-\frac{m_{L}}{\gamma_{L}})\right)+
K^{(0)}(\gamma_{L}t) \nonumber\\
&& \, \, \, \, \, \, -\gamma_{L}C_{I}
\left(-J^{(1)}(\gamma_{L}t,-\frac{m_{L}}{\gamma_{L}})
\right)+K^{(1)}(\gamma_{L}t)\Biggl)\Biggl\}, \label{j1-86}
\end{eqnarray}
where $a_{1}, b_{1}, c_{1} $ and $C_{I}, D'_{I}, F'_{I} $ are
defined in Appendix B and $J^{(0)}(a,\eta),$ $J^{(1)}(a,\eta),$
$K^{(0)}(a),$ $ K^{(1)}(a) $ in Appendix A.

Using the above mentioned formulae from Appendixes A and B (without
any additional simplifications and approximations) we get, for
example
\begin{eqnarray}
A_{K^{0}\bar{K}^{0}}(t)\nonumber&=& \frac{1+\pi }{8\pi p^{\ast}q}
\Biggl\{e^{-im_{S}t}e^{-\gamma_{S}t} \Biggl[ 1+k_{S}\Biggl]\nonumber\\
&&-e^{-im_{L}t}e^{-\gamma_{L}t}
\Biggl[1-k_{L}\Biggl]\Biggl\}+N_{K^{0}\bar{K}^{0}}(t), \label{j1-87}
\end{eqnarray}
where
\begin{eqnarray}
k_{S}=\frac{\sqrt{\gamma_{S}\gamma_{L}}}{ \gamma_{S}} \Biggl(-2 \,
i \, \gamma_{S}C_{I} +D'_{I}-F'_{I}\Biggl) , \label{j1-87a}
\end{eqnarray}
\begin{eqnarray}
k_{L}=\frac{\sqrt{\gamma_{S}\gamma_{L}}}{ \gamma_{L}} \Biggl(2 \,
i \, \gamma_{L}C_{I}-D'_{I}+F'_{I}\Biggl)  , \label{j1-87b}
\end{eqnarray}
and $N_{K^{0}\bar{K}^{0}}(t)$ is the non-oscillatory term
containing the exponential integral function $E_{i}$ and it has
following form
\begin{eqnarray}
N_{K^{0}\bar{K}^{0}}(t)\nonumber&=&\frac{1}{8\pi i p^{\ast}q}
\Biggl\{e^{-im_{S}t}e^{-\gamma_{S}t}
E_{i}(\gamma_{S}t+im_{S}t)(1+\gamma_{S} k_{S})\nonumber\\
&&+e^{-im_{L}t}e^{-\gamma_{L}t}
E_{i}(\gamma_{L}t+im_{L}t)(-1+\gamma_{L} k_{L})\nonumber\\
&&+ e^{-im_{S}t}e^{\gamma_{S}t} E_{i}(-\gamma_{S}t+im_{S}t)
\Biggl(-1+\nonumber\\
&&+\sqrt{\gamma_{S}\gamma_{L}}\Biggl[-2iC_{I}+
\frac{1}{\gamma_{S}}(-D'_{I}+F'_{I})\Biggl]\Biggl)\nonumber\\
&&+ e^{-im_{L}t}e^{\gamma_{L}t} E_{i}(-\gamma_{L}t+im_{L}t)
\Biggl(1+\nonumber\\
&&+\sqrt{\gamma_{S}\gamma_{L}}\Biggl[2iC_{I}+
\frac{1}{\gamma_{L}}(D'_{I}-F'_{I})\Biggl]\Biggl)\Biggl\}.
\label{j1-88}
\end{eqnarray}

Using the expression (\ref{j1-115}) for the derivative of $E_{i}$ (Appendix A) we can
find the derivatives which will be necessary for the following calculations, for
example
\begin{eqnarray}
\frac{\partial A_{K^{0}\bar{K}^{0}}(t)}{\partial
t}\nonumber&=&\frac{1+\pi } {8\pi
p^{\ast}q}\Biggl\{e^{-im_{S}t}e^{-\gamma_{S}t}
\Biggl(-im_{S}-\gamma_{S}(1+k_{S})\Biggl)\nonumber\\
&&+e^{-im_{L}t} e^{-\gamma_{L}t}\Biggl(im_{L}-\gamma_{L}(1+k_{L})
\Biggl)\Biggl\} \nonumber\\
&&+\Delta N_{K^{0}\bar{K}^{0}}(t), \label{j1-91}
\end{eqnarray}
where $\Delta N_{K^{0}\bar{K}^{0}}(t)$ is defined as follows
\begin{eqnarray}
\Delta N_{K^{0}\bar{K}^{0}}(t)\nonumber&=& \frac{1}{8\pi
ip^{\ast}q}\Biggl\{e^{-im_{S}t}e^{-\gamma_{S}t} E_{i}(\gamma_{S}t+im_{S}t)\Biggl(-i
m_{S}-
\gamma_{S}(1+k_{S})\Biggl)\nonumber\\
&&+e^{-im_{L}t}e^{-\gamma_{L}t} E_{i}(\gamma_{L}t+im_{L}t)\Biggl(i
m_{L}-\gamma_{L}(1+k_{L})\Biggl)\nonumber\\
&&+e^{-im_{S}t}e^{\gamma_{S}t}E_{i} (-\gamma_{S}t+im_{S}t)\Biggl(i
m_{S}-\gamma_{S}+\nonumber\\
&&+\sqrt{\gamma_{S}
\gamma_{L}}(-2i\gamma_{S}C_{I}-D'_{I}+F'_{I})\Biggl)\nonumber\\
&&+e^{-im_{L}t}e^{\gamma_{L}t}E_{i}(-\gamma_{L}t+ im_{L}t)\Biggl(-i m_{L}-
\gamma_{L}+\nonumber\\
&&+\sqrt{\gamma_{S} \gamma_{L}}(2i\gamma_{L}C_{I}+D'_{I}-F'_{I})\Biggl)\Biggl\}.
\label{j1-92}
\end{eqnarray}
There are similar expressions for $A_{\bar{K}^{0}K^{0}}(t),$
$N_{\bar{K}^{0}K^{0}}(t),$ $\frac{\partial
A_{\bar{K}^{0}K^{0}}(t)}{\partial t},$ $\Delta
N_{\bar{K}^{0}K^{0}}(t).$

The states $|K_{L}\rangle$ and $|K_{S}\rangle$ are superpositions of
$|K^{0}\rangle$ and $|\bar{K}^{0}\rangle$ ((\ref{j1-68a}),
(\ref{j1-68b})). The lifetimes of the $|K_{L}\rangle$ and
$|K_{S}\rangle$ particles may be denoted by  $\tau_{L}$ and
$\tau_{S}$, respectively, $\tau_{L}=\frac{1}{\Gamma_{L}}=(5.17\pm
0.04)\cdot 10^{-8}s$ being much longer than
$\tau_{S}=\frac{1}{\Gamma_{S}}=(0.8935\pm 0.0008)\cdot 10^{-10}s$
\cite{p:30}.

Below we calculate the difference (\ref{j1-79}) for $t\sim
\tau_{L}$
\begin{eqnarray}
h_{11}(t\sim \tau_{L})-h_{22}(t \sim \tau_{L})
=\frac{X(t\sim\tau_{L})}{Y(t\sim\tau_{L})}. \label{j1-95}
\end{eqnarray}

When we consider only the long living states $|K_{L}\rangle$ then we
may drop all the terms containing $e^{-\gamma_{S}t}|_{t\sim
\tau_{L}}$ as they are negligible in comparison with the elements
involving the factor $e^{-\gamma_{L}t}|_{t\sim \tau_{L}}.$  We also
drop all the non-oscillatory terms $N_{K^{0}K^{0}}(t),$
$N_{\bar{K}^{0}K^{0}}(t),$ $N_{K^{0}\bar{K}^{0}}(t)$ (\ref{j1-88})
present in $A_{K^{0}K^{0}}(t)$ ((\ref{j1-77})),
$A_{\bar{K}^{0}K^{0}}(t)$
 and $A_{K^{0}\bar{K}^{0}}(t)$ (\ref{j1-87}), because they are
extremally small in the region of time  $ t\sim \tau_{L}$
\cite{p:2}. Similarly, because of the properties of the
exponential integral function $E_{i},$ we can drop terms like
$\Delta N_{\bar{K}^{0}K^{0}}$ in $\frac{\partial
A_{\bar{K}^{0}K^{0}}}{\partial t}$  and   $\Delta
N_{K^{0}\bar{K}^{0}}$ (\ref{j1-92}) in $\frac{\partial
A_{K^{0}\bar{K}^{0}}}{\partial t}$ (\ref{j1-91}). This conclusion
follows from the asymptotic properties of the $E_{i}$ function
(\ref{j1-108a}) and the fact that $\Delta N_{\bar{K}^{0}K^{0}},$
$\Delta N_{K^{0}\bar{K}^{0}}$ only contain expressions
proportional to $E_{i}$.

We may now calculate the products $A_{K^{0}K^{0}}(t)\cdot
A_{\bar{K}^{0}\bar{K}^{0}}(t), $
$A_{K^{0}\bar{K}^{0}}(t)\cdot A_{\bar{K}^{0}K^{0}}(t), $\\
$\frac{\partial A_{\bar{K}^{0}K^{0}}}{\partial t}(t)\cdot
A_{K^{0}\bar{K}^{0}}(t), $  $\frac{\partial
A_{K^{0}\bar{K}^{0}}}{\partial t}(t)\cdot A_{\bar{K}^{0}K^{0}}(t)$,
that after the use of the above mentioned properties of
$N_{K^{0}K^{0}}(t)$, $\Delta N_{K^{0}K^{0}}(t)$ and performing some
algebraic transformations, lead to the following form of the
difference (\ref{j1-95}):
\begin{eqnarray}
h_{11}(t\sim \tau_{L})-h_{22}(t\sim\tau_{L}))=
\Biggl(\frac{2\pi^{2}\sqrt{\gamma_{S}\gamma_{L}}}{\pi^{2}+2\pi+1}\Biggl)\cdot
\frac{Z}{W}\neq 0, \label{j1-97a}
\end{eqnarray}
where
\begin{eqnarray}
Z= 4|p|^{2}|q|^{2}-\frac{\pi^{2}+2\pi +1
}{4\pi^{2}}\nonumber&\Biggl[1+\gamma_{S}\Biggl(4\gamma_{L}C_{I}^{2}+
\frac{1}{\gamma_{L}}(-D_{I}^{'2} -F_{I}^{'2}
+4D_{I}^{'}F_{I}^{'})\nonumber\\
&+4 i C_{I}(D_{I}^{'}-F_{I}^{'})\Biggl)\Biggl]\neq 0\label{j1-97b}
\end{eqnarray}
\begin{eqnarray}
W=2\Biggl(-C_{I}m_{L}+D_{I}^{'}-
F_{I}^{'}\Biggl)+i\Biggl[-4C_{I}\gamma_{L}+\frac{m_{L}}{\gamma_{L}}
(-D_{I}^{'}+F_{I}^{'})\Biggl]\neq 0. \label{j1-97c}
\end{eqnarray}

\section{Final remarks}
Our results lead to the conclusion that in a CPT invariant and CP
noninvariant system in the case of the exactly solvable one-pole
model, the diagonal matrix elements do not have to be equal. In the
general case the diagonal elements depend on time and their
difference, for example at $t\sim \tau_{L}$, is different from zero.
This has been clearly demonstrated in the last Section: Z and W in
(\ref{j1-97a}) are different from zero, so the difference
$(h_{11}(t)-h_{22}(t))|_{t\sim\tau_{L}}\neq 0.$ From the above
observation a conclusion of major importance can be drawn, namely
that the measurement of the difference $(h_{11}(t)-h_{22}(t)) $
should not be used for designing CPT invariance tests. This runs
counter to the general conclusions following from the Lee, Oehme and
Yang theory.

A detailed analysis of $h_{jk}(t)$, $(j,k = 1,2)$ shows that the
non-oscillatory elements $N_{\alpha ,\beta}(t), \Delta N_{\alpha
,\beta}(t)$ (where $\alpha , \beta = K^{0}, {\overline{K}}^{0}$)
is the source of the non-zero difference $(h_{11}(t) - h_{22}(t))$
in the model considered. It is not difficult to verify that
dropping all the terms of  $N_{\alpha ,\beta}(t), \Delta N_{\alpha
,\beta}(t)$ type in the formula for $(h_{11}(t) - h_{22}(t))$
gives $(h_{11}^{osc}(t) - h_{22}^{osc}(t)) = 0$, where
$h_{jj}^{osc}(t)$, $(j = 1,2)$, stands for $h_{jj}(t)$ without the
non-oscillatory terms.

To obtain the numerical estimate the real and imaginary parts of\linebreak
$(h_{11}(t\sim\tau_{L})-h_{22}(t\sim\tau_{L}))$
it is necessary to put experimentally
obtained values of $m_{S}, m_{L}, \gamma_{S}, \gamma_{L}$, etc., into
(\ref{j1-97a})-(\ref{j1-97c}). According to the literature \cite{p:6, p:8}, if the
total system is CPT--invariant but CP--noninvariant then we have (see, eg.,
\cite{p:31})
\begin{eqnarray}
p=\frac{1+\varepsilon}{\sqrt{2}} ,\ \ \ q=\frac{1-\varepsilon}{\sqrt{2}},
\label{j1-97h}
\end{eqnarray}
and hence we get
\begin{eqnarray}
\Delta_{K}=2\cdot \Re\, (\varepsilon) \label{j1-97i}
\end{eqnarray}
where $|\varepsilon| \simeq 10^{-3}$ \cite{p:6, p:8}. Putting experimental values
\cite{p:30}
\begin{eqnarray}
\Delta m = (3.489\pm 0.008)\times 10^{-12} MeV, \\
m_{S}\simeq m_{L}\simeq m_{average}=(497.648\pm 0.022) MeV \label{j1-97j}
\end{eqnarray}
and $\tau_{L},$ $\tau_{S},$ $\hbar$  into expressions (\ref{j1-97a}) - (\ref{j1-97c})
for the neutral kaon system we can obtain the following estimations
\begin{eqnarray}
\Re\, (h_{11}(t\sim\tau_{L})-h_{22}(t\sim\tau_{L})) \simeq -4.771\times 10^{-18} MeV
\label{j1-97f}
\end{eqnarray}
and
\begin{eqnarray}
\Im\, (h_{11}(t\sim\tau_{L})-h_{22}(t\sim\tau_{L})) \simeq 7.283\times 10^{-16} MeV.
\label{j1-97g}
\end{eqnarray}
So, the difference of the diagonal  matrix elements of the effective Hamiltonian for
the $K^{0}-\bar{K}^{0} $ system in one pole approximation is different from zero.
According to our evaluation
\begin{equation}
\frac{|\Re (h_{11}(t\sim\tau_{L})-h_{22}(t\sim\tau_{L}))|}{m_{average}} \equiv
\frac{|m_{K^{0}}-m_{\bar{K}^{0}}|}{m_{average}}\sim 10^{-21}. \label{j1-xx}
\end{equation}
Recent experiments give $\frac{|m_{K^{0}}-m_{\bar{K}^{0}}|}{m_{average}}\leq10^{-18}$
\cite{p:30}. So our estimation (\ref{j1-xx}) does not contradict the experimental
results.

Deviations from the LOY result estimated in \cite{p:2} have the
order of magnitude $\frac{\gamma}{m}.$ These estimations refer to
amplitudes $A_{K_{0}\bar{K}_{0}}$ and $A_{\bar{K}_{0}K_{0}}.$
However, these estimations could not be directly transformed into
the calculation of the difference $(h_{11}(\tau)-h_{22}(\tau)), $
because the difference depends not only on amplitudes of type
$A_{K_{0}\bar{K}_{0}},$ but also on their derivatives (see:
relations (\ref{j1-79}) - (\ref{j1-79b})). There are products of
type $\frac{\partial A_{\bar{K}^{0}K^{0}}(t)}{\partial
t}A_{K^{0}\bar{K}^{0}}(t)$ in the numerator of the expression
(\ref{j1-79}), whereas there aren't any derivatives in the
denominator of this expression. What is more, there is the
difference of expressions of type $\frac{\partial
A_{\bar{K}^{0}K^{0}}(t)}{\partial t}A_{K^{0}\bar{K}^{0}}(t)$ in the
numerator of (\ref{j1-79}). So, it can hardly be expected, that the
order of deviations from the LOY result of the relatively
complicated expression (\ref{j1-79}) will be the same as the order
of corrections to the LOY result of one of the following
expressions: $A_{K_{0}\bar{K}_{0}}$ and $A_{\bar{K}_{0}K_{0}}.$

If estimation (\ref{j1-xx}) is compared with a similar one obtained
in \cite{p:1}, \cite{p:33} - \cite{p:35} one can see that the
numerical value of our estimation is much larger than the value of
mentioned estimations. It is because the estimations given in the
mentioned papers were obtained using a different method for the
Lee--Fridrichs model \cite{p:15}.

The results  ($h_{11}(t)-h_{22}(t))\neq 0$ and (\ref{j1-97f}),
(\ref{j1-97g}) and (\ref{j1-xx}) seem to be very important as they
have been obtained within the exactly solvable one-pole model based
on the Breit-Wigner ansatz, i.e. the same model as used by Lee,
Oehme and Yang.

As the final remark it should also be noted that the real parts of
the diagonal matrix elements of the mass matrix $H_{\parallel},$
$h_{11}$ and $h_{22},$ are considered in the literature as masses of
unstable particles $|\textbf{1}\rangle,$ $|\textbf{2}\rangle$ (e.g.,
mesons $K_{0},$ $\bar{K}_{0}$). The interpretation of the diagonal
matrix elements of $H_{\parallel}(t=0)\equiv PHP$ is obvious (see
\cite{p:35}). They have the dimension of the energy (that is, the
mass) and $h_{jj}(0)\equiv \langle \textbf{j}|H|\textbf{j}\rangle,$
$(j=1,2).$ So their interpretation as masses of particle "1" and its
antiparticle "2" at the instant t=0 seems to be justified. Note that
 $H_{\parallel}$ has the following form
(\cite{h(t)},
\cite{p:33} - \cite{p:35})
\begin{eqnarray}
H_{\parallel}(t)=PHP+V_{\parallel}(t),
\end{eqnarray}
that is
\begin{eqnarray}
H_{\parallel}(t)\equiv H_{\parallel}(0)+V_{\parallel}(t).
\end{eqnarray}
The diagonal matrix elements of the operator $V_{\parallel}(t),$
i.e. $v_{jj}(t)=\langle j | V_{\parallel}(t) | j \rangle $, also
have the dimension of the energy and in general they depend on time.
So, the problem seems to be open: we can treat the matrix elements
of the operator $V_{\parallel}(t)$ as a time-dependent correction to
the energy or mass.

\section*{Acknowledgements}
The author wishes to thank the Referee for many critical and
inspiring remarks and Professor Krzysztof Urbanowski  for many
helpful discussions and comments on the subject treated in this
paper. The author also thanks Doctor Jaros\l{}aw Piskorski for his
help.

\appendix
\section{Appendix}
This appendix contains the relevant integrals and derivatives used
in Section 3.

Integrals $K^{(n)}(a) $ and $J^{(n)}(a,\eta) $ are defined as
follows \cite{p:2, p:28}
\begin{equation}
K^{(n)}(a)\equiv \int_{0}^{\infty}dx\frac{x^{n}}{x^{2}+1}e^{-iax},
\label{j1-98}
\end{equation}
\begin{equation}
J^{(n)}(a,\eta)\equiv
\int_{0}^{\eta}dx\frac{x^{n}}{x^{2}+1}e^{-iax}. \label{j1-99}
\end{equation}
If we assume $a \equiv (\gamma_{S/L}t) $ and $\eta \equiv
(-\frac{m_{S/L}}{\gamma_{S/L}}), $ for  $n=0 $ we get
\begin{eqnarray}
K^{(0)}(\gamma_{S/L}t)\nonumber&=&
\int_{0}^{\infty}dy\frac{1}{y^{2}+1}
e^{-i\gamma_{S/L}ty} \nonumber\\
&=&\frac{\pi}{2}e^{-\gamma_{S/L}t}-
\frac{i}{2}\Biggl[e^{-\gamma_{S/L}t}E_{i}
(\gamma_{S/L}t) \nonumber\\
&&-e^{\gamma_{S/L}t}E_{i}(-\gamma_{S/L}t)\Biggl], \label{j1-104}
\end{eqnarray}
\begin{eqnarray}
J^{(0)}(\gamma_{S/L}t,- \frac{m_{S/L}}{\gamma_{S/L}})\nonumber&=&
\int_{0}^{-\frac{m_{S/L}}{\gamma_{S/L}}}dy
\frac{1}{y^{2}+1}e^{-i\gamma_{S/L}ty} \nonumber\\
&=&-\frac{1}{2i} \Biggl[-isgn(-\frac{m_{S/L}}{\gamma_{S/L}})
e^{-\gamma_{S/L}t}\nonumber\\
&&+e^{-\gamma_{S/L}t}E_{i} \left(\gamma_{S/L}t
[1-i(-\frac{m_{S}}{\gamma_{S/L}})]\right)\nonumber\\
&&-e^{\gamma_{S/L}t}E_{i} \left(-\gamma_{S/L}t[1+
i(-\frac{m_{S}}{\gamma_{S/L}})]\right)\nonumber\\
&&-e^{-\gamma_{S/L}t}E_{i}
\left(\gamma_{S/L}t\right)\nonumber\\
&&+e^{\gamma_{S/L}t}E_{i}(-\gamma_{S/L}t)\Biggl], \label{j1-105}
\end{eqnarray}
where  $E_{i} $ is the exponential integral function and $\, sgn
\,(-\frac{m_{S/L}}{\gamma_{S/L}}) $ stands for the sign of
$(-\frac{m_{S/L}}{\gamma_{S/L}}) $.

Any other integral of  $K^{(n)}(a) $ or $J^{(n)}(a,\eta) $ for
$n>0 $ can be obtained from (\ref{j1-98}) or (\ref{j1-99}) by
differentiating (\ref{j1-98}) or (\ref{j1-99}) with respect to  $a
$ and using the Fourier identity in (\ref{j1-99}) \cite{p:2}. For
$n=1 $ we have
\begin{eqnarray}
K^{(1)}(\gamma_{S/L}t)\nonumber&=&
\int_{0}^{\infty}dy\frac{x}{y^{2}+1}
e^{-i\gamma_{S/L}ty}\nonumber\\
&=&-i\frac{\pi}{2}e^{-\gamma_{S/L}t}-
\frac{1}{2}\Biggl[e^{-\gamma_{S/L}t}E_{i}
(\gamma_{S/L}t)\nonumber\\
&&+e^{\gamma_{S/L}t} E_{i}(-\gamma_{S/L}t)\Biggl], \label{j1-106}
\end{eqnarray}
\begin{eqnarray}
J^{(1)}(\gamma_{S/L}t,-\frac{m_{S/L}} {\gamma_{S/L}})\nonumber&=&
\int_{0}^{-\frac{m_{S/L}}{\gamma_{S/L}}}
dy\frac{x}{y^{2}+1}e^{-i\gamma_{S/L}ty}\nonumber\\
&=&-\frac{1}{2} \Biggl[i sgn (-\frac{m_{S/L}}
{\gamma_{S/L}})e^{-\gamma_{S/L}t}\nonumber\\
&&-e^{-\gamma_{S/L}t}E_{i} \left(\gamma_{S/L}t[1-i(-\frac{m_{S}}
{\gamma_{S/L}})]\right)\nonumber\\
&&-e^{\gamma_{S/L}t}E_{i} \left(-\gamma_{S/L}t[1+i(-\frac{m_{S}}
{\gamma_{S/L}})]\right)\nonumber\\
&&+e^{-\gamma_{S/L}t}E_{i}
\left(\gamma_{S/L}t\right)\nonumber\\
&&+e^{\gamma_{S/L}t} E_{i}(-\gamma_{S/L}t)\Biggl]. \label{j1-107}
\end{eqnarray}
The exponential integral function $E_{i} $ is defined in the
following way \cite{p:2}, \cite{p:28}
\begin{equation}
E_{i}(\pm xy)=\pm e^{\pm xy}\int_{0}^{\infty}dt\frac{e^{-xt}}{y\mp
t},  \  \   \  \Re y>0, \ x>0. \label{j1-108}
\end{equation}
We can use the very convenient asymptotic properties of $E_{i}$
given in  \cite {p:29}
\begin{eqnarray}
\nonumber && E_{i}(0)=-\infty ,\nonumber\\
&& E_{i}(\infty )=\infty ,\nonumber \\
&& E_{i}(-\infty )=0 ,\nonumber\\
&& E_{i}(i\infty )=i\pi , \nonumber\\
&& E_{i}(-i\infty )=-i\pi.  \label{j1-108a}
\end{eqnarray}
These properties of $E_{i}$ have been used to obtain the final
result (\ref{j1-97a}) - (\ref{j1-97c}).

In our calculations we have also used the formula for the
derivative of $E_{i}$. Its final, general form is given below
\begin{equation}
\frac{dE_{i}(\pm xy)}{dx}=\frac{1}{x}e^{\pm xy}. \label{j1-115}
\end{equation}

\section{Appendix}
In this Appendix we collect from \cite {p:2} the coefficients
$a_{1}, b_{1}, c_{1} $ and $C_{I}, D'_{I}, F'_{I} $ which were used
in Section 3.

The calculations will be clearer if we write the sum of the
product \linebreak $ \sum_{\beta}
\omega_{S,\beta}^{\ast}\omega_{L,\beta} $ in the same way in which
spectral functions defined by (\ref{j1-72}), (\ref{j1-73}) were
used earlier in \cite {p:2}
\begin{eqnarray}
\nonumber&& \sum_{\beta}\omega_{S,\beta}^{\ast}\omega_{L,\beta}
\begin{array}[t]{l} \vline \, \\ \vline \,
{\scriptstyle BW} \end{array} =
\frac{\sqrt{\gamma_{S}\gamma_{L}}}{\pi
[(m-m_{S})^{2}+\gamma_{S}^{2}]
[(m-m_{L})^{2}+\gamma_{L}^{2}]}\times \nonumber\\
&&\times\{(a_{R}m^{2}+b_{R}m+c_{R})+ i(a_{I}m^{2}+b_{I}m+c_{I})\}
\label{j1-117}
\end{eqnarray}
where
\begin{eqnarray}
\nonumber&&a_{I}=I, \ \ b_{I}=(\gamma_{S}-\gamma_{L})
R-(m_{S}+m_{L})I \nonumber\\
&& c_{I}=(\gamma_{L}m_{S}-\gamma_{S}m_{L})
R+(m_{S}m_{L}+\gamma_{S}\gamma_{L}) I \label{j1-118}
\end{eqnarray}
similar formulae may be found for  $a_{R},b_{R}, c_{R}, $ where
\begin{eqnarray}
R=\frac{\Delta_{K}}{2\sqrt{\gamma_{S}\gamma_{L}}}(\gamma_{S}S+
\gamma_{L}L), \label{j1-119}
\end{eqnarray}
\begin{eqnarray}
I=\frac{\Delta_{K}}{2\sqrt{\gamma_{S}\gamma_{L}}}\frac{\Delta m
}{\gamma_{S}-\gamma_{L}}(\gamma_{S}S- \gamma_{L}L), \label{j1-120}
\end{eqnarray}
and
\begin{eqnarray}
\nonumber&& S=1+\frac{\gamma_{S}}{\pi m_{S}}+
{\cal O} ((\gamma_{S}/ m_{S})^{2}),\nonumber\\
&& L=1+\frac{\gamma_{L}}{\pi m_{L}}+{\cal O}((\gamma_{L}/
m_{L})^{2}). \label{j1-121}
\end{eqnarray}
Equations (\ref{j1-121}) result from performing the following
integration
\begin{eqnarray}
\int_{0}^{\infty}dm \;  \sum_{\alpha}|\omega_{S,\alpha }|^{2}=
\int_{0}^{\infty}dm \; \sum_{\beta}|\omega_{L,\beta }|^{2}=1.
\label{j1-122}
\end{eqnarray}
This integral follows from the initial conditions defined by
(\ref{j1-47})-(\ref{j1-49}).

This expression is now factored
\begin{eqnarray}
\nonumber&&\frac{a_{I}m^{2}+ b_{I}m+c_{I}}{[(m-m_{S})^{2}+
\gamma_{S}^{2}][(m-m_{L})^{2}+
\gamma_{L}^{2}]}= \nonumber\\
&&\frac{C_{I}m+D_{I}}{(m-m_{S})^{2}+
\gamma_{S}^{2}}+\frac{E_{I}m+F_{I}}{(m-m_{L})^{2}+ \gamma_{L}^{2}}
\label{j1-123}
\end{eqnarray}
which leads, as usual, to a linear system of equations which allows
us to calculate the coefficients $C_{I}, D_{I}, E_{I}, F_{I} $
\begin{eqnarray}
\nonumber&& E_{I}=-C_{I},\nonumber \\
&& C_{I}\Delta m+D'_{I}+F'_{I}=a_{I}\nonumber\\
&& C_{I}[(m_{L}^{2}+\gamma_{L}^{2})-(m_{S}^{2}+
\gamma_{S}^{2})]-2D'_{I}(m_{L}-2F'_{I}m_{S}=b_{I}\nonumber\\
&& D'_{I}(m_{L}^{2}+\gamma_{L}^{2})+F'_{I}(m_{S}^{2}+
\gamma_{S}^{2})+C_{I}[m_{L}(m_{S}^{2}+\gamma_{S}^{2})\nonumber\\
&&-m_{S}(m_{L}^{2}+ \gamma_{L}^{2})]=c_{I} \label{j1-124}
\end{eqnarray}
In the last formula we have introduced new definitions
\begin{eqnarray}
\nonumber&& E_{I}=-C_{I},\nonumber\\
&& D'_{I}\equiv D_{I}+C_{I}m_{L}, \ \ \ F'_{I}\equiv
F_{I}-C_{I}m_{L}. \label{j1-125}
\end{eqnarray}
\pagebreak

\end{document}